# CALCULATION OF THE FISSION-FRAGMENT YIELDS OF PRE-ACTINIDE NUCLEI ON EXAMPLE OF $^{nat}$Pb ISOTOPES

[1]V. T. Maslyuk, [1]O. A. Parlag, [1]O. I. Lendyel, [1]T. I. Marynets,
[1]M. I. Romanyuk, [2]O.S. Shevchenko, [2]Ju.Ju. Ranyuk, [2]A.M. Dovbnya

[1]Institute of Electron Physics, National Academy of Sciences of Ukraine, Uzhgorod, Ukraine
[2]National Science Center Kharkov Institute of Physics and Technology, Ukraine
Email: nuclear_photo@meta.ua

**ABSTRACT**

In the frame of proposed statistical method the calculation of fission-fragment yields (mass and charge spectra) for pre-actinide nuclei on example of $^{nat}$Pb isotopes had been conducted. The role of neutron's shells with $N = 50$ and $N = 82$ in the realizing of single- and double-humped shape of the fission-fragment yields, respectively for neutron-deficient and neutron- proficient Pb isotopes was investigated. Comparison of theoretical and experimental data of $^{nat}$Pb fission was performed with the involvement of the transformation to the ensemble of long- and short-lived nuclei fragments.

**Keywords:** Pre-actinide, nuclei, $^{nat}$Pb isotopes, fission, fragments yield, mass/charge spectrum, time transformation, 2-, 1- humped shape,

## 1. Introduction

Medium-heavy or pre-actinide nuclei with mass number $195 < A < 210$ are interesting, but relatively poorly explored objects of physics of atomic nuclei fission. In contrast to heavy nuclei with $Z^2/A \geq 36$ fissionability parameters, the fission of pre-actinide nuclei is "exotical" and requires additional excitation energy. Nevertheless, data about the peculiarities of fission that is stimulated as a rule by the interaction with high-energy nuclear particles, as well as fission-fragment yields are important for subcritical assemblies or transmutator reactors. It is well known that such objects contain the pre-actinide nuclei at the base of electron-neutron converters. These nuclei could be as a part of nuclear fuel's components or recoverable radioactive waste. On the other hand, data of nuclei fission are important for understanding the stability of atomic nuclei and the nature of nuclear forces. Although the data on the mass or charge fission-fragment yields of that nuclei currently are very limited, the existing experiments of nuclei fission, stimulated by the interaction with charged particles [1-3] or bremmstrahlung radiation [4, 5] for a wide energy range 50 MeV - 2 5 GeV should be emphasized. Theoretical studies of the peculiarities of energetic structure and medium heavy nuclei fission are significantly limited and concerns to the calculation of transient configurations and statistical characteristics of pre-actinide nuclei in the fission reactions, see [6, 7].

This work continued previous investigations [8] and is devoted to systematization of medium heavy nuclei fragments on example of Pb isotopes fission (mass and charge yields) in the frame of previously proposed statistical method [9]. Particularly, the possible asymmetry of fission fragment yields, the role of shell effects and the interpretation of experimental data where these yields have a symmetrical (one- humped) form is discussed.

## 2. Theoretical understanding

In this article the calculation of the mass/charge yields of output nucleus fission fragments with atomic mass $A_0$ and charge $Z_0$ is carried out in the frame of proposed statistical method based on the following assumptions [9]:

- The character of the mass/charge yields is determined by condition of thermodynamic ensemble's adjustment that contains clusters of fission fragments;
- The nuclei fragments fission ensemble is a constant pressure ensemble, and its thermodynamic parameters: P (pressure) and T (temperature) is determined by state of initial nucleus;
- In computing of configurational entropy the statistical non-equivalence of nucleons with different binding fraction of each fission fragments should be considered;
- The nuclear particles emission (fission neutrons, gamma radiation) maintains the constancy of the thermodynamic parameters of the nuclear fragments ensemble, namely, system's thermostat;
- During the fission neutrons' emission the internal energy of fragment nuclei changes are in process, owing to execution of work while reducing the volume of the initial nucleus at constant pressure;
- Inasmuch as atomic nuclei are small systems, it should be appreciated the fluctuations of thermodynamic parameters in the calculation of the equilibrium configurations of the nuclear fragments ensemble.

In this case the probability of realisation, for example, the *i*-th of two-nuclear fragment set is determined in the usual way via the isobaric distribution function:

$$f_i(V) = \check{S}_i \exp\{-(v_i + PV)/T\}/Z_p, \qquad (1)$$

where the statistical sum $Z_p$ is defined as $Z_p = \sum_{k,V} \check{S}_k \exp\{-(v_k + PV)/T\}$

$$v_i = \sum_{j=1,2} \cdot \sum_{\langle N_p \rangle_i} \cdot \sum_{\langle N_n \rangle_i} U_j(A_{j,i}, Z_{j,i}), \qquad (2)$$

$U_j(A_{j,i}, Z_{j,i})$ - value of the binding energy of the *j*-th nucleus fragment from the *i*-th cluster; the symbol <...> means that the summation in (1) is conducted for the *i*-th set that contains two fission fragments with the numbers of protons and neutrons that satisfy following condition:

$$\sum_{j=1,2}(N^p_{j,i} + N^n_{j,i}) + n_i) = A_0, \qquad (3)$$

where $n_i$ - number of fission neutrons, $N^p_{j,i}/N^n_{j,i}$ - number of protons/neutrons for the *j*-th fragment. The $\check{S}_i$ value - is the number of possible realizations of *i*-th set of nuclei fragments, calculated at protons/neutrons statistical nonequivalence condition of various nuclei fragments:

$$\check{S}_i = A_0!/n_i!/(\prod_{j=1,2}(N^p_{j,i}!N^n_{j,i}!)), \qquad (4)$$

The isobaric term $PV$ (see Equation (1)) was chosen in the form $PV = P(V_0 - v_0 n)$, where $V_0$ is the initial nucleus volume, $P$ is the nucleon "gas" pressure and $v_0$ is the averaged value of volume related to a single fission neutron with total number of *n*. The isobaric constant $Pv_0$ value was estimated within 4 - 5   V and was evaluated from the condition that the total fission neutron number $\bar{n}$ does not exceed 3 neutrons per fission.

Then, the distribution function $F(A_1)$ of a single fission fragment with mass $A_1$, or the same, $F(Z_1)$, with charge $Z_1$ has to be obtained by a following procedure:

- Forming the whole ensemble of post-scission fragment clusters, using for nucleons conservation conditions (1),
- The initial (not normalized) values $F(A_1)$ are obtained as the sum of probabilities of two-fragment set, containing the fission fragment with the mass $A_1$, see Equation (1). This procedure is similar to the method of histograms and must includes cumulative chains;
- the same procedure is valid for $F(Z_1)$;
- The Monte Carlo procedure must be applied to simulate the statistical fluctuations of the thermodynamical parameters of the fission fragments ensemble;
- The next step includes the normalization procedure and determination of the final values of $F(A_1)$ and $F(Z_1)$. These functions must satisfy the following normalization equations: $\sum_{<A_1>} F(A_1) = \sum_{<Z_1>} F(Z_1) = 200\%$, where $<A_1>$, $<Z_1>$ have the same meaning as in (2).

It should be noted that proposed statistical method contains no adjustable parameters, but only those that can be obtained from experiment. It provides an opportunity to investigate the features of mass and charge distributions depending on the length of the cumulative fission fragments' chain, with or without consideration of presence of short-, long-lived or stable fission fragments that are not observed by methods of semiconducting gamma spectrometry. On the other hand, this calculation method is useful for verifying of the available mass formula systematic for nuclei and for tabulation of their binding energies values.

### 3. Results and discussion

Fig. 1 shows the calculation results of the mass spectra and charge yields of fission fragments for 20 lead isotopes $^{190}$Pb – $^{210}$Pb as a basic component of $^{nat}$Pb, obtained under temperature (excitation energy) of initial nucleus with atomic mass $A_0$ and charge $Z_0$, T = 1 MeV, isobaric constant $Pv_0$ = 4.5 MeV, and no more 3 emitted neutrons per fission act was permitted. The length of cumulative chain did not exceed 10 nuclei, the values of binding energies compilation was chosen from [10]. The calculation shows a tendency toward shape changing of the yields of fission fragments from single-humped (symmetric) to double-humped dependencies when passing from neutron-deficient (as at A = 190) to neutron-proficient Pb isotopes, which occurs in the vicinity of lead isotopes $^{202-204}$Pb. Note that this tendency is more pronounced for the mass spectra of fission fragments and lowest values of temperature T.

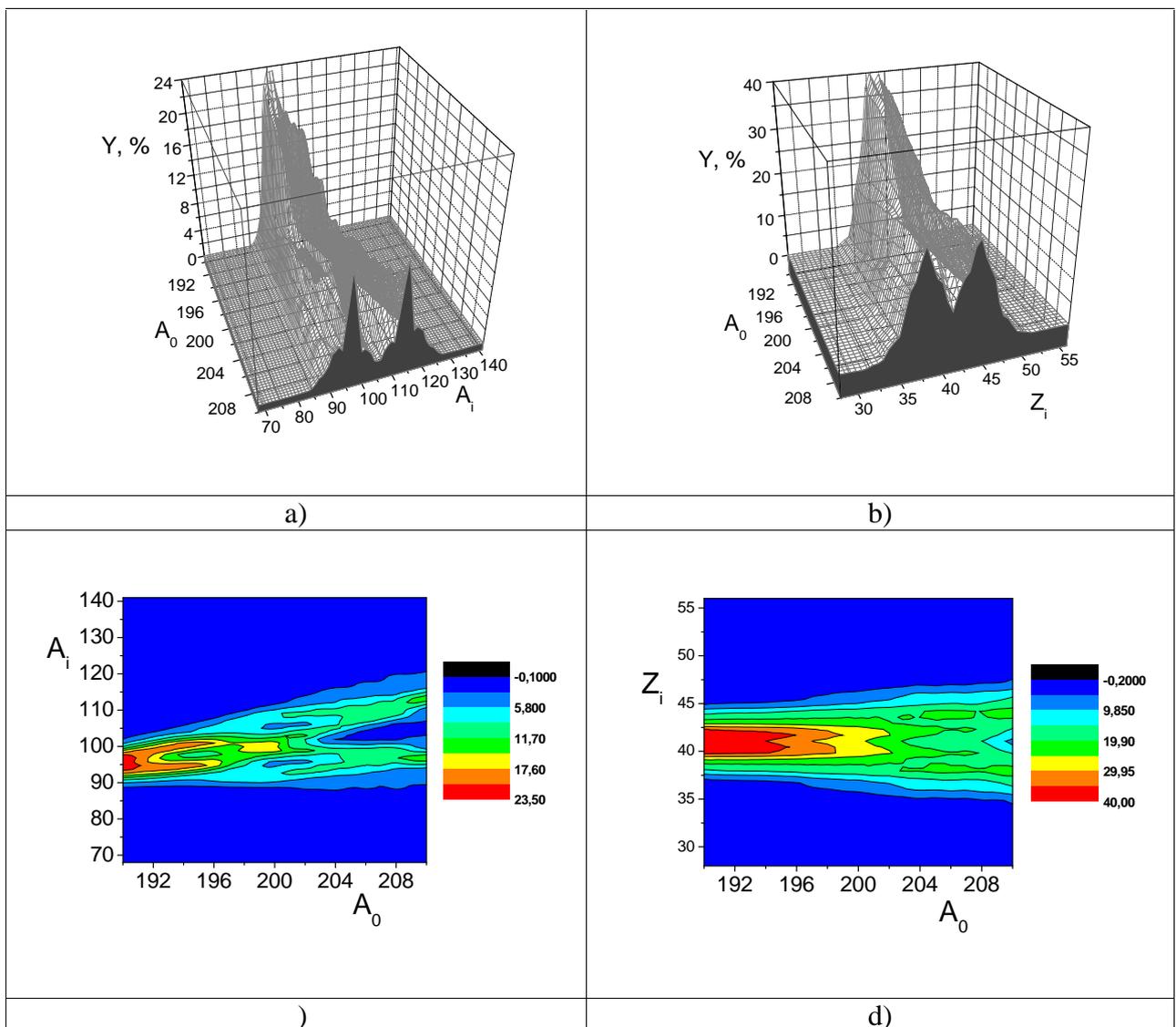

Fig.1. (**Color online**) The values of mass - a), c), and charge, - c), d) spectra of lead isotopes' fission fragments are presented in the form of 3D (a, b) and contour (c, d) diagrams.

It is interesting to investigate the reasons of the isotopes fission-fragment yields' asymmetry that formed $^{nat}$Pb, and role of 50 and 82 nuclear filled shells in this process. The calculation shows that the entropy term S = ln ($Š_i$), where $Š_i$ defined in (4) is responsible for the symmetrization of the fission fragment yields with the rise of the nuclear temperature. This is due to the fact that $S$ reaches a maximum if $N_{1,i}^p = N_{2,i}^p$, $N_{1,i}^n = N_{12i}^n$ and role of this component is important at high temperatures, T. However, for neutron-deficient nuclei (e.g. $^{190}$Pb), the mentioned symmetrization effect of fission fragment yields is enhanced by contribution of fission fragment yields that reside near to the filled neutron shells at N = 50, as the $^{95-97}$Zr, $^{96,97}$Mo. In case of $^{210}$Pb fission, for example, the double-humped form of fission fragments is caused by presence in fission fragments the neutron-proficient isotopes $^{89,91,93}$Kr, $^{95,96,97}$Sr and $^{115-117}$Pd, $^{110-113}$Ru with neutron numbers close to non-filled shell N = 82, those role is more significant than contribution of entropic term S, responsible for the symmetric fission of nucleus.

In other words, the asymmetry or the double-humped nature of mass and charge spectra (as follows from calculations) are caused as a result of the competition of unfilled neutron shells either N = 50 or N = 82 under formation of heavy and light fission fragments, respectively, for $^{nat}$Pb base isotopes. The same reason is crucial at pre-actinide of the mass/charge fission fragments spectra for other pre-actinide fission nuclei, as Ta, Re, In, Au and Hg.

The experimental results of $^{nat}$Pb photofission confirmed the single-humped form of nuclear fission-fragment yields distribution, and there is a tendency of peak expanding of gamma–quantum energy: from 60 MeV to 2.5 GeV [4, 5]. Anisotropy, particularly, double-hamped form of $^{nat}$Pb isotopes fission-fragment yields was not observed in this experiment and the cause is to be sought in the conditions of experiment. Thus, the $^{nat}$Pb fission fragments yields measurements was carried out only after 0.5 - 1 hour after irradiation, the gamma-active nuclei was identifying only, short-lived isotopes was disregarded. These facts significantly affected the results of investigations.

In this respect it is interesting to investigate the properties of isotopic fission fragments spectra for initial nuclei from $^{190}$Pb to $^{210}$Pb. Figure 2 shows the results of such calculations, for example, for Ru isotope yields, had been obtained for the same calculated parameters as in Figure 1.

For understanding of the role of statistical fluctuations that are typical for small systems, the isotopic spectra (on Figure 2) were obtained by using the Monte Carlo procedure. The fluctuation range was up to 20% for all and $P€_0$ and 100 statistical tests were curried out. Obviously, that maxima of Ru isotopes yields (peaks in Figure 2) was displaced from stability field to unstable in β⁻ transforming nuclei when passing from neutron-deficient ($^{190}$Pb) to neutron- proficient ($^{210}$Pb) isotopes. This tendency is much more expressed for the light nuclear fragments. Specified regularities on Figure 2 are typical for the other isotopes of $^{nat}$Pb fission fragments ensemble.

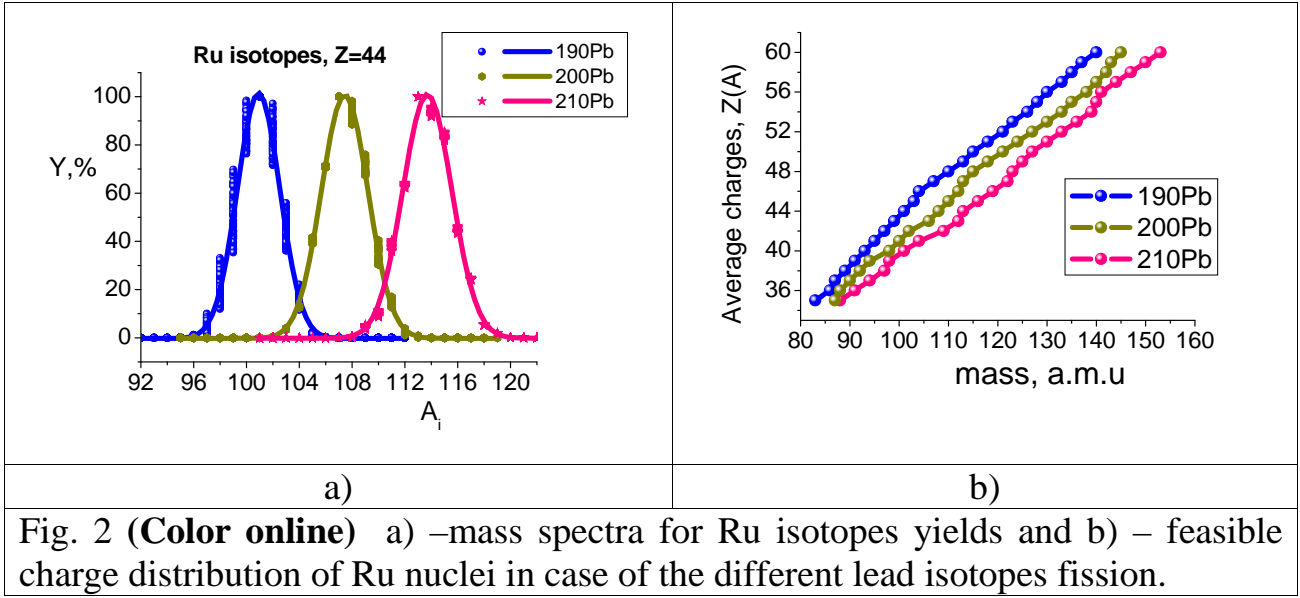

Fig. 2 **(Color online)**  a) –mass spectra for Ru isotopes yields and b) – feasible charge distribution of Ru nuclei in case of the different lead isotopes fission.

It allows to suggest a model of the fission fragments yield shapes' evolution of $^{nat}$Pb isotopes over time, that is recorded in measurements with the $\Delta t_m$ delay. Understandably that result of such investigations depends on the ratio of long- and short-lived nuclei in the ensemble of fission fragments. For example, in case of $^{208}$Pb isotope the fission fragments mass spectrum shape can vary from double-humped (Figure 1), when during the experiment we can observe long- and short-lived fission fragments, that are possible at $\Delta t_m \leq 1$ sec; to the single-humped one, when $\Delta t_m \leq 1$ hour and long-lived nuclei fragments are fixed only. Last case meets the experimental conditions [4] and its results could be explained by the dynamics of nuclear reactions. Thus, the isotopes $^{103,105}$Ru had been experimentally observed in maximum of fission fragments yields for $^{208}$Pb [4, 5], could appear as result of the β⁻ transformations chains:

$$^{103,105}\text{Nb} \rightarrow {}^{103,105}\text{Mo} \rightarrow {}^{103,105}\text{Tc} \rightarrow {}^{103,105}\text{Ru},$$

That occur during ~ 10 min., that is less than in experimental conditions, where $\Delta t_m$ ~ 0.5 - 1 hour. Reduction of double-humped mass and charge spectra structure (Figure 1), can arise even during less time. Considering that maximum of the light-fission fragment yields are formed by the Rb isotopes during 1 sec owing to chain of β⁻ transformation according to the scheme:

$$^{97}\text{Rb} \rightarrow {}^{97}\text{Sr} \rightarrow {}^{97}\text{Y} \rightarrow {}^{97}\text{Zr},$$

the long-living isotopes of Zr are formed, which was observed in [4].

The results of such isotopic transformations are schematically illustrated in Fig. 3, which presents the evolution of the mass spectra shape, for example, for $^{208}$Pb isotope fragment fission yields versus the possible β⁻ - decay chain length.

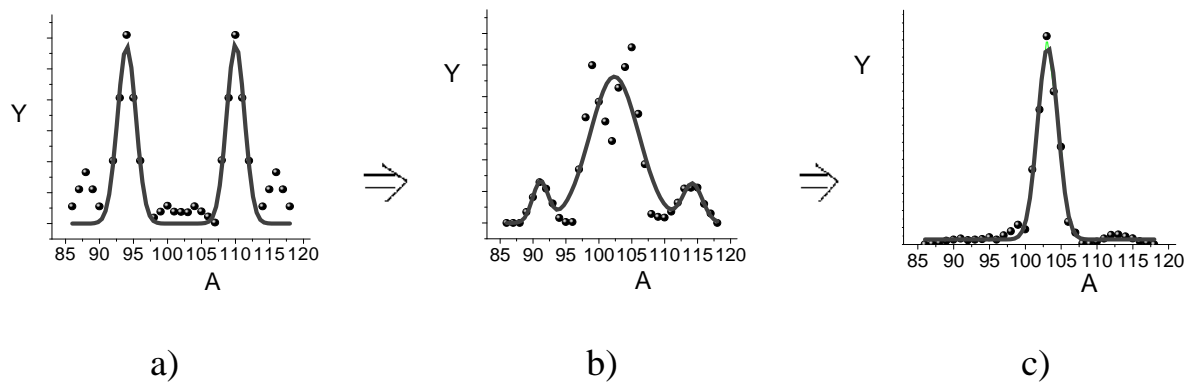

a)                  b)                  c)

Fig. 3 Change of shape of the mass fission fragments yields of the $^{208}$Pb had been obtained without, - (a) and considering the possibility of emission 3, - (b) or 5, - (c) β⁻ particles during the conversions of isotopes.

Summing up, the research results allow us to expect the yield and peculiarities of the pre-actinide materials' fission products transformations that are important for nuclear-power engineering, especially, for subcritical assemblies' reactors.

The analysis shows that for a wide number of pre-actinide nuclei the competition of unfilled neutron shells $N = 50$ and $N = 82$ is important and determines the peculiarities of fission-fragment yields from asymmetric (double-humped) to single-humped dependence.

## 4. Acknowledgements